\documentclass[prab,aastex,amsmath,amssymb,twocolumn,floatfix,aps,showpacs]{revtex4-2}

\usepackage{graphicx}
\usepackage{amsmath}
\usepackage{amssymb}
\usepackage{esint}
\usepackage{textcomp}
\usepackage{xcolor}
\definecolor{myblue}{rgb}{0.0, 0.0, 0.6}
\usepackage{hyperref}
\hypersetup{
  colorlinks = true,
  citecolor  = myblue,
  linkcolor  = myblue,
  urlcolor   = myblue
}

\usepackage{lineno}
\usepackage{setspace}

\begin{document}

\title{
  Comments on the paper \\
  ``Eliminating beam-induced depolarizing effects in the hydrogen jet target
  for high-precision proton beam polarimetry at the Electron-Ion Collider \cite{Rathmann:2025jgp}''
}

\author{A.~A.~Poblaguev}
\email{poblaguev@bnl.gov}
\affiliation{Brookhaven National Laboratory, Upton, New York 11973, USA}

\date{\today}

\begin{abstract}
A critical review of the methodology used in F.\,Rathmann \emph{et al.}, Phys. Rev. Accel. Beams 29, 021001 (2026), to evaluate beam-induced depolarization of the Atomic Polarized Hydrogen Gas Jet (HJET) target at the Electron--Ion Collider (EIC) is presented. It is shown that several key assumptions underlying that analysis—including the introduction of a photon emission threshold, the application of Fermi’s Golden Rule to coherent hyperfine transitions, the interpretation of power broadening as a physical linewidth increase, and the treatment of spatial magnetic fields—are either incorrect or internally inconsistent. As a consequence, the predicted large depolarization effects are demonstrated to be artifacts of the adopted methodology rather than genuine physical phenomena. A consistent quantum-mechanical treatment based on the time-dependent Schr\"odinger equation shows that beam-induced depolarization probabilities at the EIC are negligibly small.
\end{abstract}

\maketitle

For more than two decades, the Atomic Polarized Hydrogen Gas Jet Target (HJET) \cite{Zelenski:2005mz,Poblaguev:2020qbw} has been successfully used in the RHIC Spin Program for absolute calibration of the proton beam polarization. However, operation of the HJET at the Electron--Ion Collider (EIC) requires additional consideration owing to the significantly more stringent requirements on systematic uncertainties in beam polarization measurements, as well as the approximately threefold increase in the average beam current and the nearly order-of-magnitude reduction in both bunch spacing and bunch length.

An evaluation of possible beam-induced depolarization \cite{HERMES:1998twm} of the jet target at the EIC was initiated in Ref.\,\cite{Rathmann:2025jgp}, where it was claimed that the jet target would be unavoidably depolarized at the EIC if operated at the 120\,mT holding field used at RHIC.

In the present Comment, we argue that this conclusion is not supported by a consistent quantum-mechanical analysis.

Beam-induced depolarization may arise from the time-dependent magnetic field generated by the bunched proton beam. Assuming a Gaussian longitudinal bunch density, the temporal dependence of the induced magnetic field at a radius $r$ from the beam axis can be approximated as
\begin{equation}
  B_\text{ind}(r,t) = B_\text{pk}(r)\sum_{k=-\infty}^{\infty}
  \exp\!\left[-\frac{(t-k\tau_b)^2}{2\sigma_t^2}\right],
  \label{eq:B(r,t)}
\end{equation}
where for the EIC flattop $\sigma_t=0.2\,\text{ns}$, $\tau_b=11.027\,\text{ns}$, and the peak value of the bunch-induced field $B_\text{pk}(r)$ reaches its maximum $B_m=3.03\,\text{mT}$ at $r_m=1.04\,\text{mm}$ \cite{Rathmann:2025jgp}.

Since $B_\text{ind}$ is periodic in time, it can be expanded into harmonic components:
\begin{align}
  B_\text{ind}(r,t) &= \frac{\sqrt{2\pi}\sigma_t}{\tau_b}B_m F_B(r)
  \nonumber \\ &\times
  \left[
    1 + \sum_{\eta=1}^{\infty}
    2\,e^{-(\omega_\eta \sigma_t)^2/2}
    \cos(\omega_\eta t)
  \right],
  \label{eq:B_harm}
\end{align}
where $\omega_\eta = 2\pi \eta f_b$, $f_b=\tau_b^{-1}=90.683\,\text{MHz}$, and the radial dependence $F_B(r)=F(r)/F(r_m)\le1$ is expressed in terms of $F(r)$ defined in Eq.~(45) of \cite{Rathmann:2025jgp}. The normalization factor $\sqrt{2\pi}\sigma_t/\tau_b$ ensures that the zeroth harmonic equals the time-averaged value of $B_\text{ind}(r,t)$. The equivalence of the two representations of $B_\text{ind}(r,t)$, \eqref{eq:B(r,t)} and \eqref{eq:B_harm}, can be readily verified by direct numerical calculation.

The amplitude of the $\eta$-th harmonic is therefore
\begin{equation}
  B_1^{(\eta)}(r)
  = B_m F_B(r)e^{-2\pi^2 \eta^2\sigma_t^2 f_b^2}
  \le 0.136\,\text{mT}.
  \label{eq:B1}
\end{equation}
Since all harmonics are coherent for $t=0$,
\begin{equation}
  B_1^{(0)}(r) + \sum_{\eta=1}^{\infty}{2B_1^{(\eta)}(r)} = B_m F_B(r) \le B_m.
  \label{eq:harmSum}
\end{equation}

It is well known (see, e.g., Ref.\,\cite{Rathmann:2025jgp}) that, in a static holding field ${\boldsymbol B}_\text{hold}$, the ground state of the hydrogen atom is described by a four-level hyperfine system with eigenstates $|i\rangle$ and eigenvalues $E_i$ of the time-independent Hamiltonian $H_0(B_\text{hold})$.

In the presence of a time-dependent perturbation $H_1(t)$, the atomic wave
function may be written \cite{LandauQM} as
\begin{equation}
  \Psi(t) = \sum_k a_k(t)\,|k\rangle e^{-iE_k t/\hbar},
\end{equation}
and the Schr\"odinger equation leads to
\begin{equation}
  i\hbar\frac{d a_k(t)}{dt}
  = \sum_n {\cal M}_{kn}(t)\,a_n(t)\,e^{i\omega_{kn} t},
  \label{eq:an(t)}
\end{equation}
where ${\cal M}_{kn}(t)=\langle k|H_1(t)|n\rangle$ and
$\omega_{kn}=(E_k-E_n)/\hbar$.

For a single harmonic $\omega=2\pi \eta f_b$ from
Eq.\,\eqref{eq:B_harm},
\begin{equation}
  {\cal M}_{kn}(t)=\hbar\omega_R^{(kn)}\cos(\omega t), \quad
  \omega_R^{(kn)} = -2\mu_e B_1^{(\eta)} S_{kn},
\end{equation}
where $\mu_e=-14.012\,\text{MHz/mT}$ is the electron magnetic
moment.

The values of the spin-dependent factors $S_{kn}$ in ${\cal M}_{kn}(t)$ \cite{Beijers:2005, Poblaguev:2026jpv} depend on whether the oscillating field is parallel ($\sigma$) or perpendicular ($\pi$) to ${\boldsymbol B}_\text{hold}$, and on the mixing angle $\theta$, defined by $\tan(2\theta)=B_c/B_\text{hold}$, where $B_c=50.7\,\text{mT}$ \cite{Rathmann:2025jgp},
\begin{equation}
  S_{kn}^\pi =
  \begin{pmatrix}
    0             & \sin\theta    &  0             &  \cos\theta    \\
    \sin\theta    & 0             &  \cos\theta    &  0             \\
    0             & \cos\theta    &  0             & -\sin\theta    \\
    \cos\theta    & 0             & -\sin\theta    &  0
  \end{pmatrix},
\end{equation}

\begin{equation}
  S_{kn}^\sigma =
  \begin{pmatrix}
    1   &  0               &  0 &  0               \\
    0   &  \cos{2\theta}   &  0 & -\sin{2\theta}   \\
    0   &  0               & -1 &  0               \\
    0   & -\sin{2\theta}   &  0 & -\cos{2\theta} 
  \end{pmatrix}.
\end{equation}

The physical role of the diagonal terms in $S_{ij}^\sigma$ is transparent.
For a stationary state $|i\rangle$, the wave function evolves as
$\exp(-iE_i t/\hbar)$. The longitudinal component of the beam-induced magnetic
field modifies the energy $E_i$ and thus alters the phase evolution.
At the matrix-element level, this effect is encoded in the diagonal elements
of $S_{ij}^\sigma$, which represent field-induced shifts of the stationary
state energies.

To analyze the behavior of the system in the vicinity of a resonant transition, consider a two-level subsystem $\{ij\}$ initially prepared in state $|i\rangle$, with $a_i(0)=1$ and $a_j(0)=0$. For small detuning,
$|\Delta\omega| = |\omega-\omega_{ij}| \ll |\omega_{ij}|$, and for short interaction times $t \ll |\Delta\omega|^{-1}$ (but still much larger than $|2\omega_{ij}|^{-1}$), one finds
\begin{equation}
  a_j(t) = -i\frac{\omega_R}{2}
  \int_0^t dt'\left(
    e^{-i\Delta\omega t'} + e^{-i2\omega_{ij}t'}
  \right)
  \approx -i\frac{\omega_R t}{2}.
  \label{eq:coherence}
\end{equation}

For longer interaction times, both transitions
$|i\rangle \leftrightarrow |j\rangle$  must be treated consistently, which leads to the well-known result~\cite{Rabi:1937dgo}
\begin{equation}
  a_j(t) = -i\,\frac{\omega_R}{\Omega}
  \sin\!\left(\frac{\Omega t}{2}\right), 
  \qquad
  \Omega = \sqrt{\omega_R^2 + \Delta\omega^2}.
  \label{eq:detune}
\end{equation}
This expression is fully consistent with Eq.~\eqref{eq:coherence} in the limit of short times $t \ll 1/\Omega$.

The corresponding transition probability is
\begin{equation}
  \Pi_{ij} = |a_j(t)|^2
  = \frac{\omega_R^2}{\Omega^2}
  \sin^2\!\left(\frac{\Omega t}{2}\right),
  \label{eq:detun}
\end{equation}
which, for nonzero detuning, may appear as an effective broadening of the transition line. However, this effect arises purely from coherent interference and does not correspond to any physical smearing or modification of the atomic energy levels themselves.

For specified matrix elements and initial conditions $a_n(0)$,
Eq.\,\eqref{eq:an(t)} can be solved exactly, for example, by numerical integration. In Ref.\,\cite{Poblaguev:2026jpv} it was shown that, for EIC flattop operation and the potentially resonant transitions $\{ij\}=\{13\},\,\{14\},\,\{23\},\,\{24\}$,
$\langle\omega_R^{\{ij\}}(t)\rangle\tau_\text{int}/2\ll1$ (see Eq.\,\eqref{eq:coherence}) due to the exponential suppression~\eqref{eq:B1} of $B_1^{(\eta)}$ at high harmonic numbers and the small total resonant interaction time $\tau_\text{int}$. Consequently, the corresponding matrix elements can be neglected, ${\cal M}_{ij}\to0$, on the right-hand side of Eq.\,\eqref{eq:an(t)}, and the four-level system reduces to two independent subsystems,
$\{12\}$ and $\{34\}$, which considerably simplifies the integration.
Including the time dependence of $\omega_R(t)$ and the detuning $\Delta\omega(t)$ along the atomic trajectory, it was found that the beam-induced depolarization at the EIC flattop is negligible, $\lesssim10^{-4}$.

\medskip

The methodology used in Ref.\,\cite{Rathmann:2025jgp} does not account for the dynamical evolution described by Eq.\,\eqref{eq:an(t)}. Instead, an atom-track-averaged Rabi frequency $f_R=\omega_R/2\pi$ was employed.

Although Eq.\,\eqref{eq:B_harm} can be obtained by inverse Fourier transformation directly from Eq.\,(19) of Ref.\,\cite{Rathmann:2025jgp}, it was not explicitly used in that analysis. Moreover, the oscillation amplitudes (and hence the Rabi frequencies) were incorrectly evaluated there.

For example, using Eq.\,(37) of Ref.\,\cite{Rathmann:2025jgp} and $L_\text{int}=30.6\,\text{mm}$, one can readily evaluate the track-averaged oscillation amplitude as
\begin{equation}
  \langle B_1^{(\eta)}\rangle \approx 0.18\,B_1^{(\eta)}
  \le 25\,\mu\text{T}.
\end{equation}
However, Fig.\,16 of Ref.\,\cite{Rathmann:2025jgp} shows
$\langle B_1^{(0)}\rangle\approx1500\,\mu\text{T}$.
Furthermore, if the reported 6th harmonic amplitude of
$1174\,\mu\text{T}$ were correct, the sum of only the first seven harmonics, according to Eq.\,\eqref{eq:harmSum}, would exceed $15.3\,\text{mT}$, already far larger than the peak field of $3.03\,\text{mT}$.

Thus, the Rabi frequency was overestimated by approximately a factor of 60, and the depolarization probability by roughly a factor of 3600 when approximation~\eqref{eq:coherence} is applicable.

\medskip

In Ref.~\cite{Rathmann:2025jgp}, the enhanced occurrence of resonant conditions at the EIC flattop is attributed mainly to two effects:
\emph{(i)} ``power broadening'' due to Rabi spin precession, and
\emph{(ii)} the dense spacing of the resonant frequencies (as a function of $B_\text{hold}$) for the $\{24\}$ transitions, accompanied by a beam-magnetic-field-induced shift of the hydrogen energy levels,\footnote{
Both effects, Rabi spin precession and the beam-magnetic-field-induced shift of the energy levels, are intrinsic properties of the evolution equations~\eqref{eq:an(t)}. Therefore, no separate or standalone treatment of these effects is required in an analysis based directly on Eq.~\eqref{eq:an(t)}.}.

According to Ref.~\cite{Rathmann:2025jgp}, the total power-broadened linewidth $\sigma_f^\text{total} \approx 2.38\,\text{MHz}$—inferred from the Rabi frequency corresponding to a $200\,\mathrm{\mu T}$ oscillating field (with unspecified frequency)—implies that
\begin{quote}
  ``additional transitions within 2.38~MHz of any harmonic frequency could exhibit resonant behavior.''
\end{quote}
  
Even setting aside the interference-based interpretation~\eqref{eq:coherence} of the apparent ``broadening,'' several independent arguments invalidate the assumption that the hyperfine linewidths undergo any genuine physical broadening under the conditions considered:
\\ \noindent$\bullet$\quad
The assumption of  ``additional transitions within 2.38\,MHz of any harmonic frequency'' contradicts the superposition principle. Each harmonic component of the driving field acts independently, and off-resonant components cannot induce resonant transitions.
\\ \noindent$\bullet$\quad
The \emph{Rabi frequency} is not a frequency in the dynamical sense. Rather, it represents a conveniently normalized amplitude of the oscillating field. Consequently, $f_\text{Rabi}$, as defined in Eq.~(34) of~\cite{Rathmann:2025jgp}, is unrelated to either the oscillation frequency of the field or the beam bunch frequency. Therefore, for harmonic components not explicitly included in the evaluation of $f_\text{Rabi}$, the resonant behavior cannot depend on the value of $f_\text{Rabi}$ calculated in this manner.
\\ \noindent$\bullet$\quad
If $\sigma_f^\text{power}$ were to describe a genuine broadening of the hyperfine energy levels, all atomic states would decay to the lowest-energy state $F=0$ on a time scale of order $1/\sigma_f^\text{power}$, even in the absence of the proton beam (but in the presence of a $200\,\mathrm{\mu T}$ monochromatic oscillating magnetic field of arbitrary frequency). Such behavior is clearly unphysical and contradicts experimental observations.

\medskip

It was claimed in Ref.~\cite{Rathmann:2025jgp}:
\begin{quote}
  ``A key innovation introduced in this study is the formulation of beam-induced depolarization in terms of a photon emission threshold: a cutoff frequency $f_\text{cut}$, above which the likelihood of resonant transitions is significantly reduced due to the steep falloff in spectral power.''
\end{quote}

The photon emission rate $\dot{N}^{\,\text{avg}}_{\gamma,\text{total}}$
was defined following Eqs.~(27)--(29) and (37)--(39) of~\cite{Rathmann:2025jgp}, and the depolarization threshold $f_\text{cut}$ was introduced in relation to the instantaneous number of hydrogen atoms $N_\text{atoms}$ in the interaction region.

Although the total photon rate as a function of $f$ is \emph{mainly} proportional to the depolarization probability $\Pi(f)$, it cannot be used as a proxy for $\Pi(f)$ because it is not properly calibrated. Specifically:
\\ \noindent$\bullet$\quad
No properties of the hydrogen hyperfine states were included in the calculation of the photon emission rate $\dot{N}^{\,\text{avg}}_{\gamma,\text{total}}$.
\\ \noindent$\bullet$\quad
The proportionality between $\dot{N}^{\,\text{avg}}_{\gamma,\text{total}}$ and the depolarization probability depends explicitly on the choice of $L_\text{int}$, which renders the calibration of the photon emission rate ambiguous and unreliable.
\\ \noindent$\bullet$\quad
For any given hydrogen atom in the jet, the depolarization probability cannot depend on the jet radius $r_\text{at.\,beam}$ (see Eq.~(29) of~\cite{Rathmann:2025jgp}).

Moreover, for a low-density hydrogen gas jet, the depolarization probability of an individual atom cannot depend on the total number of hydrogen atoms in the jet. Therefore, if $\dot{N}^{\,\text{avg}}_{\gamma,\text{total}}$ is interpreted as a measure of the depolarization probability, the cutoff frequency $f_\text{cut}$ effectively becomes an arbitrarily chosen parameter. Consequently, the distributions of $\dot{N}^{\,\text{avg}}_{\gamma,\text{total}}(t)$ shown in Figs.~11 and~16 of Ref.~\cite{Rathmann:2025jgp} cannot be used for a quantitative evaluation of beam-induced depolarization. In particular, the harmonic cutoff $n_\text{cut}=55.8$ derived in Fig.~16 does not possess predictive power, and conclusions based on this value cannot be regarded as reliable.

Nevertheless, $\dot{N}^{\,\text{avg}}_{\gamma,\text{total}}$ may still be used for approximate comparisons of depolarization levels in different experiments. However, the same comparison can be obtained in a simpler and more transparent manner \cite{Steffens:2018gzz,Lenisa:2020bxj} using Eq.~\eqref{eq:B_harm}. For example, using the RHIC-related value $f_\text{cut}^\text{RHIC}=551.4\,\text{MHz}$ (independent of its physical interpretation), one readily finds the corresponding value for the EIC flattop,
\begin{equation}
  f_\text{cut}^\text{EIC}
  = \frac{\sigma_t^\text{RHIC}}{\sigma_t^\text{EIC}}\,
    f_\text{cut}^\text{RHIC}
  = 5059.1\,\text{MHz}.
\end{equation}
This result can be compared with the value 5061.1\,MHz reported in Table~III of~\cite{Rathmann:2025jgp}.

\medskip

Another claim in Ref.~\cite{Rathmann:2025jgp} states:
\begin{quote}
  ``To validate this photon emission framework, a rigorous quantum mechanical analysis using proper Breit---Rabi matrix elements and stimulated transition rates was performed.''
\end{quote}

To support this statement, Fermi’s Golden Rule was applied in Appendix~C to evaluate the beam-induced transition probabilities. However, several questions arise regarding the validity of the equations used in Ref.~\cite{Rathmann:2025jgp}:

\noindent$\bullet$\quad
Equation~(C2) contains an additional factor $V_\text{int}$, described as
``the effective interaction volume swept out by the atomic beam.'' 
Since Eq.~(C2) does not include any dependence on the hydrogen atom density distribution, the factor $V_\text{int}$ should not appear in this expression. In particular, the standard form of Fermi’s Golden Rule~\cite{Bethe:1964IQM} contains no factor associated with an interaction volume.

\noindent$\bullet$\quad
The dependence of the transition probability on the beam-induced magnetic photon density is already fully incorporated into the perturbative Hamiltonian $H_1(t)$. Therefore, the additional factor $S(f)$ should be omitted from Eq.~(C2).

\noindent$\bullet$\quad
The explicit time dependence of the perturbation,
\[
  H_1(t) = \hbar \omega_R \cos(2\pi f t),
\]
was omitted in the transition from Eq.~(C2) to Eqs.~(C3) and (C4).

\noindent$\bullet$\quad
As a consequence, the resonance dependence of the transition rates $\Gamma_{ij}(f)$---analogous to that in Eq.~\eqref{eq:detun}---is lost.

In the particular case where the oscillating magnetic field is restricted to the direction parallel to the static holding field, only the $\{24\}$ transition is allowed. In this situation, the four-level system reduces to an effective two-level subsystem $\{24\}$ governed by the well-known Rabi formula~\eqref{eq:detun}, in which the transition probability is an oscillatory function of time. This example clearly demonstrates that the probabilities presented in Appendix~C of Ref.~\cite{Rathmann:2025jgp} cannot adequately describe hyperfine transitions in hydrogen induced by an external oscillating magnetic field.

Thus, Fermi’s Golden Rule, \emph{as applied in Ref.~\cite{Rathmann:2025jgp}}, is not applicable to beam-induced depolarization effects in the hydrogen jet target, and the conclusions presented in Appendix~C should therefore be disregarded.

\medskip

The conclusion of Ref.~\cite{Rathmann:2025jgp} that jet target depolarization is unavoidable in EIC measurements with
$H_\text{hold}\sim120\,\text{mT}$ is, in fact, supported by only a single quantitative estimate.
In the example discussed in Sec.~VI.D, the $\{24\}$ transition
(with frequency $f_{24}=3627\,\text{MHz}$) was evaluated using a
``spatial field distribution with an effective rf field amplitude of about
$B_1=2.0\,\text{mT}$.''  
Within this approximation, the Rabi frequency is
$\Omega=6.9\times10^7\,\text{rad/s}$
(as calculated in Ref.~\cite{Rathmann:2025jgp}),
leading to the conclusion that the affected atoms undergo significant depolarization.

However, independently of the interpretation of the claimed ``broadening,''
a stimulated resonance transition $\{24\}$ requires an external magnetic photon frequency $f\approx f_{24}$. The density of such photons decreases exponentially with frequency, as shown in Eq.~\eqref{eq:B1}. Therefore, the appropriate estimate of the Rabi frequency is
\begin{align}
  \omega_R^{(24)} &\le
  \frac{2|\mu_e|}{\hbar}
  \sqrt{2\pi}\,\sigma_t f_b B_m
  \exp\!\left(-2\pi^2 f_{24}^2 \sigma_t^2\right)
  \sin{2\theta}
  \nonumber \\
  &\approx 290\,\text{rad/s},
\end{align}
which results in a negligible transition probability,
\begin{equation}
  \Pi_{24} \le 1.9\times10^{-8}.
\end{equation}

If the spatial magnetic field is nevertheless used to evaluate the
$\{24\}$ transition probability, its oscillation frequency $f_b$
must be explicitly taken into account. This introduces a large detuning
$\Delta\omega = 2\pi\,(f_{24}-f_b)$,
which, according to Eq.~\eqref{eq:detune}, leads to strong suppression of the transition probability and, consequently, to negligible depolarization.

\medskip

To highlight several other methodological issues in Ref.~\cite{Rathmann:2025jgp}, we reconsider the $\{24\}$ transition example, strictly following the procedure suggested in Ref.~\cite{Rathmann:2025jgp}.

Using Eq.~(34) of Ref.~\cite{Rathmann:2025jgp} and the numerical values provided for $B_1=2\,\text{mT}$ and $\sin{2\theta}=0.392$, we obtain
$\omega_R^\text{sp} = 1.38\times10^8\,\text{rad/s}$.
It is worth noting that $\omega_R^\text{sp}$ is a factor of two larger than the corresponding $\Omega$ reported in Ref.~\cite{Rathmann:2025jgp}. Consequently, using $\tau_\text{int}=0.94\,\mathrm{\mu s}$,
\begin{equation}
  \Pi_{24}^\text{\cite{Rathmann:2025jgp}}
  = \sin^2\!\left(\frac{\omega_R^\text{sp}\tau_\text{int}}{2}\right)
  = 72\%.
\end{equation}

However, even a small variation in the atomic velocity
($1807\,\text{m/s}$) and, consequently, in $\tau_\text{int}$ leads to drastic changes in the calculated transition probability:
\begin{align}
  \tau_\text{int}\approx0.955\,\rm{\mu s} \quad&\Rightarrow\quad \Pi_{24}=\phantom{10}0\%, \\
  \tau_\text{int}\approx0.932\,\rm{\mu s} \quad&\Rightarrow\quad \Pi_{24}=100\%.
\end{align}
This extreme sensitivity is explained by the large value
$\omega_R^\text{sp}\tau_\text{int}/2\approx65\gg\pi$.

For a correct interpretation of such estimates, the transition probability should be written as
\begin{equation}
  \Pi_{24}
  = \sin^2\!\left(\frac{\omega_R^\text{sp}\tau_\text{int}}{2}\right)
  = \frac{1-\cos\!\left(\omega_R^\text{sp}\tau_\text{int}\right)}{2}.
\end{equation}
When $\omega_R^\text{sp}\tau_\text{int}\gg\pi$, unavoidable atom-to-atom variations of
$\omega_R^\text{sp}$ and $\tau_\text{int}$ lead to an average transition probability
$\Pi_{24}^\text{eff}=0.5$, corresponding to complete depolarization of the considered atom.

Within the framework employed in Ref.~\cite{Rathmann:2025jgp}, the only means of suppressing the $\{24\}$ resonance transition is to increase the holding field $B_\text{hold}$, thereby reducing $\sin 2\theta$. Consequently, achieving a depolarization level below $1\%$ ($\omega_R^\text{sp}\tau_\text{int}/2\sim0.1$) for the $\{24\}$ transition requires
\begin{equation}
 B_\text{hold}^\text{1\%}
\approx B_c\left(%
 \sin{2\theta}\,\frac{0.1}{\omega_R^\text{sp}\tau_\text{int}/2}
 \right)^{-1}
 \approx84\,\text{T},
\end{equation}
which is far beyond practical limits and would effectively preclude the use of an HJET at the EIC.

It may also be noted that, if one applies exactly the same reasoning used to infer large depolarization due to the $\{24\}$ transition at the EIC to the RHIC case, one would be forced to conclude that unavoidable depolarization of the HJET target should also occur at RHIC (regardless of the exact value of $B_\text{hold}\sim120\,\text{mT}$). However, no such effect has been observed during many years of HJET operation.

\medskip

To summarize, the depolarization analysis presented in Ref.~\cite{Rathmann:2025jgp} relies on incorrect assumptions and misinterpretations of several relevant physical effects. As a result, it leads to misleading predictions for beam-induced transition probabilities in the hydrogen gas-jet target at the EIC.

Importantly, when the correct harmonic amplitudes~\eqref{eq:B1}---which are in exact agreement with Eq.~(19) of \cite{Rathmann:2025jgp}---are used, the formalism of Ref.~\cite{Rathmann:2025jgp} itself leads to the conclusion that the jet target depolarization at the EIC is negligibly small.

Some mistakes in Ref.~\cite{Rathmann:2025jgp} that are not directly related to the calculation of beam-induced depolarization are discussed in the technote~\cite{osti_3382181}.

\section*{Acknowledgments}
This work is authored by an employee of Brookhaven Science Associates, LLC under Contract No.\,DE-SC0012704 with the U.S. Department of Energy.

\bibliographystyle{apsrev4-2}
%

\end{document}